\documentclass{aa} 
\usepackage{amsmath,amssymb,amsfonts}
\usepackage{graphicx}
\usepackage{txfonts}
\usepackage{enumerate} 
\usepackage{colordvi} 
\usepackage{bm}
\usepackage{epsfig}
\usepackage{hyperref}
\usepackage[dvipsnames]{xcolor}
\usepackage{color, colortbl}
\usepackage[T1]{fontenc} 
\usepackage{braket}
\usepackage{euclid-Like}
\usepackage{placeins}
\usepackage{tabularx}
\usepackage{comment}
\usepackage{multirow}
\usepackage{csquotes}
\graphicspath{{./Figures/}}
\usepackage{glossaries}
\usepackage{orcidlink}

\let\Gamma\varGamma
\let\Delta\varDelta
\let\Theta\varTheta
\let\Xi\varXi
\let\Pi\varPi
\let\Sigma\varSigma
\let\Upsilon\varUpsilon
\let\Phi\varPhi
\let\Psi\varPsi
\let\Omega\varOmega

\definecolor{amaranth}{rgb}{0.9, 0.17, 0.31}
\definecolor{forestgreen(web)}{rgb}{0.13, 0.55, 0.13}
\definecolor{lavender(web)}{rgb}{0.9, 0.9, 0.98}
\definecolor{cosmiclatte}{rgb}{1.0, 0.97, 0.91}
\definecolor{jonquil}{rgb}{0.98, 0.85, 0.37}
\definecolor{khaki(x11)(lightkhaki)}{rgb}{0.94, 0.9, 0.55}
\definecolor{thistle}{rgb}{0.85, 0.75, 0.85}

\newcommand{\lcdm}{\ensuremath{\varLambda\mathrm{CDM}}}

\newcommand{\de}{\mathrm{d}}

\newcommand{\aperpo}{a_{\perp 0}}

\newcommand{\Hperp}{H_{\perp}}
\newcommand{\Hperpo}{H_{\perp 0}}
\newcommand{\Hpar}{H_{\parallel}}

\def\be{\begin{equation}}
\def\ee{\end{equation}}
\def\ba{\begin{eqnarray}}
\def\ea{\end{eqnarray}}
\def\eqi{\begin{equation}}
\def\eqf{\end{equation}}
\def\eqia{\begin{eqnarray}}
\def\eqfa{\end{eqnarray}}
\def\lcdm{$\Lambda$CDM }
\def\lltb{$\Lambda$LTB }

\def\d {\mathrm{d}}
\definecolor{valecol}{rgb}{0,0.5, 1.}
\definecolor{linkblue}{rgb}{0,0,0.8}
\definecolor{linkgreen}{rgb}{0,0.5,0}
\hypersetup{linkcolor=linkblue, citecolor=linkgreen, urlcolor=linkblue}

\defcitealias{Blanchard:2019oqi}{EC19}

\usepackage[nameinlink,noabbrev]{cleveref}
\Crefname{equation}{Eq.}{Eqs.}
\Crefname{section}{Sect.}{Sects.}
\Crefname{figure}{Fig.}{Figs.}
\defcitealias{IST:paper1}{EC20}

\begin{document}

\title{Constraining $\sf\Lambda$LTB models with galaxy cluster counts from next-generation surveys}

\newcommand{\orcid}[1]{\orcidlink{#1}}
\author{Z. Sakr \orcid{0000-0002-4823-3757}\thanks{\email{zsakr@irap.omp.eu}}$^{1,2,3}$, A.~Carvalho$^{4,5}$, A.~Da Silva\orcid{0000-0002-6385-1609}$^{4,5}$, J.~Garc\'ia-Bellido\orcid{0000-0002-9370-8360}$^{6}$,  J.P.~Mimoso\orcid{0000-0002-9758-3366}$^{4,5}$, D.~Camarena\orcid{0000-0001-7165-0439}$^{7}$, S.~Nesseris\orcid{0000-0002-0567-0324}$^{6}$, C.J.A.P.~Martins\orcid{0000-0002-4886-9261}$^{8,9}$, N. Nunes$^{4,5}$\orcid{0000-0002-3837-6914}, and D. Sapone\orcid{0000-0001-7089-4503}$^{10}$}

\institute{$^{1}$ Institut de Recherche en Astrophysique et Plan\'etologie (IRAP), Universit\'e de Toulouse, CNRS, UPS, CNES, 14 Av. Edouard Belin, 31400 Toulouse, France.\\
$^{2}$ Institut f\"ur Theoretische Physik, University of Heidelberg, Philosophenweg 16, 69120 Heidelberg, Germany.\\
$^{3}$ Universit\'e St Joseph; Faculty of Sciences, Beirut, Lebanon.\\
$^{4}$ Instituto de Astrof\'isica e Ci\^encias do Espa\c{c}o, Faculdade de Ci\^encias, Universidade de Lisboa, Campo Grande, PT-1749-016 Lisboa, Portugal.\\
$^{5}$ Departamento de F\'isica, Faculdade de Ci\^encias, Universidade de Lisboa, Edif\'icio C8, Campo Grande, PT1749-016 Lisboa, Portugal.\\
$^{6}$ Instituto de F\'isica Te\'orica UAM-CSIC, Campus de Cantoblanco, E-28049 Madrid, Spain.\\
$^{7}$ Department of Physics and Astronomy, University of New Mexico, Albuquerque, New Mexico 87106, USA.\\
$^{8}$ Centro de Astrof\'{\i}sica da Universidade do Porto, Rua das Estrelas, 4150-762 Porto, Portugal.\\
$^{9}$ Instituto de Astrof\'{\i}sica e Ci\^encias do Espa\c co, Universidade do Porto, Rua das Estrelas, 4150-762 Porto, Portugal.\\
$^{10}$ Departamento de Física, FCFM, Universidad de Chile, Blanco Encalada 2008, Santiago, Chile.\\
}

\date{\today}

\authorrunning{Z. Sakr et al.}

\titlerunning{$\sf\Lambda$LTB model from galaxy cluster counts.}


\abstract
{The Universe's assumed homogeneity and isotropy is known as the cosmological principle. It is one of the assumptions that led to the Friedmann-Lema\^{\i}tre-Robertson-Walker (FLRW) metric and is a cornerstone of modern cosmology, because the metric plays a crucial role in the determination of the cosmological observables. Thus, it is of paramount importance to question this principle and perform observational tests that may falsify it.}
{Here, we explore the use of galaxy cluster counts as a probe of a large-scale inhomogeneity, which is a novel approach to the study of inhomogeneous models, and we determine the precision with which future galaxy cluster surveys will be able to test the cosmological principle.}
{We present forecast constraints on the inhomogeneous Lema\^{\i}tre-Tolman-Bondi (LTB) model with a cosmological constant and cold dark matter, basically a $\Lambda$CDM model endowed with a spherical, large-scale inhomogeneity, from a combination of simulated data according to a compilation of `Stage-IV' galaxy surveys. For that, we followed a methodology that involves the use of a mass function correction from numerical $N$-body simulations of an LTB cosmology.
}
{When considering the \lcdm fiducial model as a baseline for constructing our mock catalogs, we find that our combination of the forthcoming cluster surveys will improve the constraints on the cosmological principle parameters and the FLRW parameters by about $50\%$ with respect to previous similar forecasts performed using geometrical and linear growth of structure probes, with $\pm20\%$ of variations depending on the level of knowledge of systematic effects.
}
{These results indicate that galaxy cluster abundances are sensitive probes of inhomogeneity and that 
next-generation galaxy cluster surveys will thoroughly test homogeneity at cosmological scales, tightening the constraints on possible violations of the cosmological principle in the framework of $\Lambda$LTB scenarios.
}

\keywords{Cosmology: observations -- Galaxies: clusters: general -- cosmological parameters --Space vehicles: instruments -- Surveys -- Methods: statistical -- Methods: data analysis}

   \maketitle

\section{Introduction} \label{sec:intro}
The standard cosmological model is based on the assumption that space-time can be well described by the Friedmann-Lema\^{\i}tre-Robertson-Walker (FLRW) metric, which stems from a generalization of the Copernican Principle, stating that our location is not special in the Universe, and also from 
observations suggesting that, on sufficiently large scales, on the order of $100\,\mathrm{Mpc}$, the Universe is homogeneous and isotropic \citep{Maddox:1990yw,2dFGRS:2005yhx,Saadeh:2016sak}.

One of the tests of this principle could be achieved by assuming an alternative metric, such as the inhomogeneous metric of the Lemaître-Tolman-Bondi model (LTB), and confronting it with current observations \citep{Garcia-Bellido:2008vdn,February:2009pv,Valkenburg:2012td,Redlich:2014gga,Camarena:2021mjr}, studying its phenomenology through simulations \citep{2012PDU.....1...24A,2022A&A...664A.179M}, or forecasting on the ability of forthcoming surveys to constrain such a deviation from the standard assumption \citep{ReviewDoc,Euclid:2022ucc}. 
In this work, we undertake this task and attempt to determine the precision achievable on the cosmological parameters of the $\Lambda$LTB model (i.e.,\ the LTB model endowed with a cosmological constant) by using galaxy cluster abundance data obtained from forthcoming surveys. This follows from previous studies, which show that cluster counts are capable of placing constraints 
on the standard cosmological model and its extensions \citep{Rapetti:2009ri, Ilic:2019pwq}.

The study of clusters has had a major role in establishing today's most accepted cosmological model, the cosmological constant and cold dark matter ($\Lambda$CDM) model, by pinning down the values of dark matter and dark energy abundances \citep{SPT:2018njh}. Clusters' progenitors are the ingredients of our Universe, associated with the primordial inflationary stage, after which the growth of the density perturbations was amplified by gravity, creating what we now know as large-scale structures (LSSs) \citep{Peebles:1982ff}. As such, the abundance of galaxy clusters is very sensitive to the LSS power spectrum and to the growth rate of their density perturbations. 
On top of that, galaxy cluster abundances are also sensitive both to the cosmological parameters and to the late evolution of the Universe, as these objects were the last to form according to the bottom-up hierarchical formation model \citep{Blumenthal:1984bp}. We need, however, to properly calibrate the mass observable proxy used to relate the mass function, which describes the mass distribution of dark matter halos, to the observed clusters. Therefore, a miscalibration could be one of the sources of the discrepancy found in the value of the matter amplitude fluctuation parameter, $\sigma_8$, when constrained by deep probes in comparison to local ones \citep{Sakr:2021jya}. Thus, we discuss later in this study the modeling of the $\Lambda$LTB mass function, as well as the calibration of the mass proxy accessed through a scaling relation proper to each survey.

Forecasts on constraints on the $\Lambda$LTB model were previously presented in \citet{ReviewDoc}, and recently in \cite{Euclid:2022ucc} using cosmic microwave background anisotropies \citep{Planck:2018vyg}, mock data of baryonic acoustic oscillations \citep{Blanchard:2019oqi}, the Compton-$y$ distortion \citep{Fixsen:1996nj}, and kinetic Sunyaev-Zeldovich data \citep{Reichardt:2020jrr}. In this work, we perform a similar study, but aim to benefit from the nonlinear scales that the galaxy clusters additionally probe, expected to be obtained from constructed “Stage-IV”-like mock catalogs, such as \Euclid \citep[][]{Laureijs11,Scaramella-EP1}. The latter is an M-class space mission of the European Space Agency (ESA) which will allow the optical detection of clusters previously unattainable in terms of the depth and area covered. Along with it, we also consider clusters detected from the Legacy Survey of Space and Time (LSST) survey, performed at the \textit{Vera C. Rubin} Observatory (and hereafter named \textit{Rubin}) \citep{Abell:2009aa} and the eROSITA survey X-ray based cluster counts \citep{2012MNRAS.422...44P}, and the Dark Energy Spectroscopic Instrument (DESI) \citep{DESI2016}. The former two, using richness as the same mass observable proxy, are complementary in terms of redshift and sky coverage, while eROSITA and DESI will allow us, each with a different detection method -- respectively, X-ray and spectroscopy -- to further extend the redshift range as well as the number of clusters detected. 

Our paper is organized as follows: in \Cref{sec:models} we briefly review the modeling of galaxy cluster counts in $\Lambda$LTB. In \Cref{sec:probes} we present the surveys' specifications and scaling relations modeling. Lastly, our results are presented and discussed in \Cref{sec:results} before we conclude in \Cref{sec:conclusions}.

\section{Modeling galaxy cluster counts in $\sf\Lambda$LTB}  \label{sec:models}
\subsection{The $\Lambda$LTB model}\label{Sec:LLTB}

The $\Lambda$LTB model is a spherically symmetric, inhomogeneous space-time with a cosmological constant, $\varLambda$, and cold dark matter (CDM), as in \citet{2014MNRAS.438L...6V} or \citet{Euclid:2022ucc}. 
For the sake of self-consistency, we now summarize the key equations of the model, from the aforementioned papers, which are necessary to understand the galaxy cluster counts methodology detailed in \Cref{Sec:LLTB_halomass}.  

The $\Lambda$LTB line element is of the form
\begin{equation}
    ds^2 = -c^2 {\rm d}t^2 + \frac{R'^2(t,r)}{1-K(r)\,r^2}\, {\rm d}r^2 + R^2(t,r)\,\Big( {\rm d}\theta^2 + \sin^2 \theta \; {\rm d}\phi^2 \Big)\,,
    \label{eq:metric}
\end{equation}
where $c$ is the speed of light, $K(r)$ is an arbitrary function of the spatial coordinate, $r$, and a prime denotes a partial derivative with respect to $r$. If $K$ is constant and $R(t,r) = a(t)\,r$, Eq.~\eqref{eq:metric} becomes the FLRW line element where $a(t)$ is the FLRW scale factor. From Eq.~\eqref{eq:metric} one can define longitudinal $a_\parallel = R'(t,r)$ and transverse $a_\perp = R(t,r)/r$, scale factors, and longitudinal, $\Hpar$, and transverse, $\Hperp$, expansion rates given by
\begin{equation}
H_\parallel \equiv \frac{\dot{a_\parallel}}{a_\parallel},
\qquad
H_\perp \equiv \frac{\dot{a}_\perp}{a_\perp},
\label{eq:expansion_rates}
\end{equation}
where the dots represent partial derivatives with respect to cosmic time. Assuming a cosmological constant, the general relativity (GR) equations allow one to derive a FLRW-like equation for the $\Lambda$LTB model,  
\begin{equation}
\frac{H^2_\perp}{H^2_{\perp 0}} = \Omega_{\rm m,0} \,\Bigg( \frac{a_{\perp0}}{a_\perp} \Bigg)^3 + \Omega_{K,0}\,\Bigg(\frac{a_{\perp0}}{a_\perp} \Bigg)^2 + \Omega_{\varLambda,0}\, ,
\label{eq:friedmann}
\end{equation}
where $H_{\perp0} \equiv \Hperp(t_0,r)$, $a_{\perp0} \equiv a_\perp(t_0,r)$, and $\Omega_{\varLambda,0}$, $\Omega_{K,0}$, 
$\Omega_{\rm m,0}$ are the present-day, $t_0$, 
density parameters associated with the cosmological constant, curvature, and matter, respectively. These parameters  
are functions of the coordinate $r$ and satisfy the closure relation $\Omega_{\rm m,0}(r)+ \Omega_{K,0}(r)+\Omega_{\varLambda,0}(r)=1$.
Using the GR geodesics equation with the metric Eq.~\eqref{eq:metric} gives    
\begin{align} 
\frac{\d t}{\d z} &= - \frac{1}{(1+z)\,H_\parallel (t,r)} \, , \label{eq:geodesics_1}\\
\frac{\d r}{\d z} &= - \frac{c\, \sqrt{1- K(r)\,r^2}}{(1+z)\,a_\parallel(t,r)\,H_\parallel(t,r)}\,.
\label{eq:geodesics_2}
\end{align}
As is seen, Eq.~\eqref{eq:geodesics_1} and Eq.~\eqref{eq:geodesics_2} are essential in our analysis because their integration provides a relation between $r$, $t$, and $z$, allowing the $\Lambda$LTB galaxy cluster's mass function to be expressed as a function of redshift, which is a key quantity in the modeling of the survey cluster counts described in \Cref{sec:probes}.


As is shown in \citet{Euclid:2022ucc}, the \lltb model can be 
parameterized by the arbitrary functions $K(r)$ and $m(r)$ that enter in the definition of 
\begin{equation}
\label{eq:Omegas}
\Omega_{\rm m,0}(r) = \frac{2\,G\,m(r)}{\Hperpo^2 \,\aperpo^3 \,r^3}\,,\,\, 
\quad 
\,\, \Omega_{K,0}(r) = -\frac{K(r)\: c^2}{\Hperpo^2\,\aperpo^2}\,, 
\end{equation}
where $m(r)$ is the Euclidean mass profile of the large-scale inhomogeneity
\begin{equation}
    m(r) = \int^r_0 {\rm d} r' \, 4\pi \, \rho_{\rm m}(t,r') \, a_\parallel \, a^2_\perp \, r'^2\,,
    \label{eq:m_r}
\end{equation} 
and where $\rho_{\rm m}(t,r)$ is the matter density. 
The cosmological constant density parameter is $\Omega_{\varLambda,0}(r) = \varLambda \,c^2/(3\,\Hperpo^2)=1-\Omega_{\rm m,0}(r)-\Omega_{K,0}(r)$. 
Once the radial coordinate, $r$, is fixed in such a way that $m(r) \propto r^3$, with a normalization described below, the curvature profile, $K(r)$, is the sole free function left in the model, 
which can be modeled as 
\begin{equation}
K(r)= K_{\rm B} + \left(K_{\rm C} - K_{\rm B}\right) \, P_3 (r/ r_\sfont{B} ) \,, \label{eq:kr}
\end{equation}
where $r_\sfont{B}$ is the comoving radius of the spherical inhomogeneity, $K_{\rm B}$ is the curvature outside the inhomogeneity, and $K_{\rm C}$ is the central inhomogeneity curvature. The function
$P_3$ enables us to specify at convenience the radial profile of the central void. We modeled this function, as in \citet{Euclid:2022ucc}, as
\begin{align}
P_{3}(x)= \left\{\begin{array}{ll}
1 - \exp \big[-(1-x)^3/x] & \mbox{ for }  0  \le x < 1\\
0 & \mbox{ for } 1 \leq x 
\end{array}\right. \,, 
\end{align}
with $x=r/ r_\sfont{B}$. 
From the density parameters in Eq.~\eqref{eq:Omegas} and the above closure relation, we determined the missing normalization of the mass, 
$m(r) = \Omega_{\rm m,0}^{\rm{\, out}}\,(H_{0}^{\rm{\, out}})^2\,r^3/(2\,G)$, as well as $K_{\rm B} = - \Omega_{K,0}^{\rm{\, out}}\,(H_{0}^{\rm{\, out}}/c)^2$, 
which allowed us to determine the volume-averaged integrated matter density contrast, 
\begin{align}
\delta(r,t_0)& 
 = \frac{3\, m(r)}{4\pi \, R^3(t_0,r) \, \rho_{\rm m}^{\rm{\, out}}(t_0)} - 1 = \nonumber  \,\\ 
& = \frac{\Omega_{\rm m,0}(r)}{\Omega_{\rm m,0}^{\rm{\, out}}}\, \left[ \frac{H_{\perp 0}(r)}{H_{\perp 0}^{\rm{\, out}}} \right]^2 -1 \,. 
\label{deltar}
\end{align}
In these expressions the quantities denoted by the superscript ``out'' are taken at $r=r_{\rm B}$. 
We note that the curvature profile outside the LTB void ($r\ge r_{\rm B}$) is set to $K(r)=K_{\rm B}$, and the $K_{\rm C}$ is directly related to the central under-density contrast  $\delta(r=0,t_0)$, as is discussed in \citet{Euclid:2022ucc}.

\subsection{Halo mass function in the $\sf\Lambda$LTB model.} \label{Sec:LLTB_halomass}
The formalism adopted in this work is based on 
the mass function derived from LTB $N$-body simulations in 
\citet{2012PDU.....1...24A}. Although the latter deals with the original LTB model, we assume that this strategy is still valid for $\Lambda$LTB if the dark energy is accounted for. Indeed, simulations in FRLW metrics show that in general the nonlinear LSS formation in $\Lambda$CDM is obtained in models that change the background, by replacing the growth in the new theory \cite{Baldi:2012ky}, a fix in line with what we adopted next in our modeling.

In the context of the $\Lambda$LTB model, we need to account for the density perturbations at cluster scales inside and outside the $\Lambda$LTB void. 
The variance of the density perturbations at very large radii, outside the void, matches that of a FLRW cosmology: 
\begin{equation}
        \sigma^2(M,z) = \frac{1}{2\pi^2} \int_0^\infty \de k\,k^2\,P_{\rm m}(k,z)\,W^2\left[\textcolor{black}{k\,R_{\rm th}(M)}\right]\,,
\label{eq:msigma}
\end{equation}
where $M$ is the mass inside a sphere of radius $R_{\rm th}(M)= \left[3\,M\,/\,(4\,\pi\,\rho_{\rm m})\right]^{1/3}$, $\rho_{\rm m}$ is the matter density, $W(k\,R_{\rm th})$ is the Fourier transform of a top-hat filter at the scale $R_{\rm th}$, and $P_{\rm m}(k,z)$ is the linear matter power spectrum of a $\Lambda$CDM universe at redshift $z$ and wave number $k$. 

Following \citet{2012PDU.....1...24A}, inside the $\Lambda$LTB void, $\sigma(M,z)$ needs to be scaled by a factor, $f$, which is the ratio between density perturbations at two regimes:
\begin{equation}
    f(t,r) = \frac{\delta_\alpha(t,r)}{\delta_\alpha (t, r \rightarrow \infty)}\, ,    
\end{equation}
where $\delta_\alpha$ is the linear density perturbation, with $\delta_\alpha (t, r \rightarrow \infty)$ matching the linear density contrast of the $\Lambda$CDM model.
For small values of shear (see Eq.~\eqref{eq:epsilon} below), it can be approximated by
\begin{equation}
    \delta_\alpha(t,r) = \bar \delta (t,r)\, \left[1 + \alpha\,\epsilon(t,r) \right]\,,    
    \label{eq:delta_alpha}
\end{equation}
which depends on the normalized shear parameter, $\epsilon$. The latter is defined as 
\begin{equation}
\epsilon \equiv \sqrt{\frac{2}{3}\frac{\Sigma^2}{\Theta^2}}=\frac{H_\perp-H_\parallel}{2\,H_\perp+H_\parallel}\, ,
\label{eq:epsilon}
\end{equation}
giving the ratio between the square of the background shear, $\Sigma$, and the expansion parameter, $\Theta$, where $\Sigma^2=\Sigma_{ij}\,\Sigma^{ij}$ is the square of the background shear,
and $\Theta$ the expansion parameter of a congruence of comoving geodesics (see \cite{2009JCAP...09..028G} for further details). %
Finally, in Eq.~\eqref{eq:delta_alpha}, $\alpha$ is a correction parameter that can be constrained from simulations or observations that we will discuss in \Cref{sec:results}, and $\bar \delta(t,r)$ is the local linear density contrast given by 
\begin{equation}\label{eq:pert_sol_0}
\bar \delta(t,r)\propto\,D(\Omega_{\rm m,0}(r),R(t,r)/R_0(r))\,,
\end{equation}
where
\begin{align}
 D(\Omega_{\rm m,0}(r), a)&=a \, .\, { }_2F_1
 \left[\frac{1}{3};1;\frac{11}{6};\frac{\Omega_{\rm m,0}(r)-1}{\Omega_{\rm m,0}(r)}\,a\right]\,
,\end{align}
with $_2F_1(a,b;c;z)$ the Gauss hypergeometric function and $D$ the growth factor in a $\Lambda$CDM universe with matter parameter $\Omega_{\rm m,0}$ \citep{Buenobelloso:2011sja,Nesseris:2015fqa}. We note that in our case we observe the redshift, therefore all the above equations were computed in $z(r,t)$ using Eq.~\eqref{eq:geodesics_1} and Eq.~\eqref{eq:geodesics_2} to then interpolate between the variables.\\
In this framework, \citet{2012PDU.....1...24A} modeled the mass function in LTB based on a FLRW mass function, with $\sigma$ replaced by $\sigma(M,z)\,f(t,r)$. Here, we adopted the same method and wrote the comoving number density of halos of mass, $M$, and redshift, $z$, as
\begin{equation}
    n(M,z) = - \frac{\rho_{{\rm m},0}}{M}\,\mathcal{F}\big[\sigma(M,z)\,f(t,r)
\big] \,  \frac{\d \ln \big( \, [\sigma(M,z)\,f(t,r) \big ]^{-1} \, \big)}{\d M} \,,
\label{eq:mf}
\end{equation}
where $\rho_{{\rm m},0}=\rho_{{\rm m}}(z=0)$, and the $\mathcal{ F}$ function taken from \cite{2016MNRAS.456.2486D} is
\begin{equation}
    \mathcal{ F}(\sigma) =\nu \, A\sqrt{\frac{2\,a}{\pi}}\left[1+\left(\frac{1}{a\,\nu^2}\right)^p\right]\  \ \exp\left[-\frac{a\,\nu^2}{2}\right]\,.
\end{equation}
$\nu\equiv\delta_{\rm c}/\sigma(M,z)$, where $\delta_{\rm c}$ is the linear density threshold at the present time outside the void for nonlinear collapse and $(a,\, A,\, p)=(0.3295,\, 0.7689,\, 0.2536)$. We rescaled following \cite{2016MNRAS.456.2486D} to match the survey mass definition. 

We modified the \texttt{monteLLTB} code  
used in \cite{Camarena:2021mjr} to include a cluster counts module that implements the halo mass function recipe described above.\footnote{\url{https://github.com/davidcato/monteLLTB}} The linear mater power spectrum was normalized using the $\sigma_{8,0}$ parameter, defined as the root mean square of the variance of density fluctuations at $R=8\, h^{-1}\,\mathrm{Mpc}$ in $\Lambda$CDM. In this work, we provide forecasts involving the following cosmological parameters: $h=H_0/(100\,\text{km\,s$^{-1}$\,M$^{-1}$\,pc})$ (where $H_0:=\Hperp(r\rightarrow \infty)=\Hpar(r\rightarrow \infty)$ is the Hubble constant),
$\Omega_{\rm m,0}:=\Omega_{\rm m,0}(r\rightarrow \infty)$, $\sigma_{8,0}$ (hereafter denoted as $\sigma_8$), $\delta_0:=\delta(r=0,t_0)$, and $z_{\rm B}$ is the boundary redshift corresponding to $r_{\rm B}$. The last two parameters are specific to the LTB void model, whereas the first three parameters of this list are in common with the $\Lambda$CDM cosmological model. 
Regarding the later, we adopted a flat $\Lambda$CDM fiducial model with $h=0.67$, $\Omega_{\rm m,0}:=\Omega_{\rm m,0}^{\rm \, out}=0.32$, $\Omega_{K,0}:=\Omega_{K,0}^{\rm \, out}=0$ (i.e.,\ $K_{\rm B}=0$), $\sigma_8=0.83$, and a spectral index of primordial scalar perturbations, $n_{\rm s}=0.965$.

\section{Survey specifications and scaling relations modeling} \label{sec:probes}
\subsection{\Euclid and \textit{Rubin} surveys' richness-based cluster counts }\label{Sec:model_Euclid_LSST}

\subsubsection{\Euclid-like cluster number counts forecast}

To estimate the number of cluster counts, we used the forecast of \cite{2016MNRAS.459.1764S}, in which the observable quantity is the observed mass, $M_{\rm obs}$, with a sky coverage of $\Omega_{\rm tot} = 15\,000\,{\rm deg}^2$. The estimated number counts for a redshift bin, $z_l$, and mass bin, $M_{{\rm obs},m}$, can be expressed as
\begin{align}
     N_{l, m} = \int_{\Omega_{\rm tot}} \d \Omega &\int^{z_{l+1}}_{z_l}\d z \;\frac{\d V}{\d z\,\d\Omega}\,   
    \notag \\[2mm]
     & \times
     \int^{+\infty}_0 \d M \;n(M,z) \frac{1}{2} \left[ \text{erfc}  (x_m) - \text{erfc}(x_{m+1})\right]\,,
\end{align}
where $n(M,z)$ is the mass function, $\d \Omega$ is the solid angle element in units of steradian, erfc is the complementary error function, and
$\frac{\d V}{\d z\,\d\Omega}$ is the derivative of the comoving volume with respect to the redshift and solid angle element, defined as
\begin{equation}
 \frac{\d V}{\d z\,\d\Omega}= \, \frac{(1+z)^2\, d_{\rm A}(z)^2 c}{\Hpar(z)}, 
 \label{eq:volume_element}
\end{equation}
where $d_{\rm A}(z)=R[t(z),r(z)]$ is the angular diameter distance and $c$ the speed of light. Finally, $x_m=x\,(M_{{\rm obs}, m, \ell})$ is defined in each mass bin, $m$, and redshift bin, $\ell$, \footnote{we drop from now on the redshift index from the scaling relations when there is a $z$ dependence for better readability.} as
\begin{equation}
    x\,(M_{{\rm obs},m}) = \frac{    \ln (M_{{\rm obs}, m}/M_0) - \ln (M/M_0) -\ln (M_\mathrm{bias}/M_0)}{ \sqrt{2 \, \sigma^2_{ M_{{\rm obs}, m}}}}\,,
\end{equation}
where $M_0 = 1 \, h^{-1}\, M_\odot $.
We defined the mass bias and the variance, $\sigma^2_{M_{{\rm obs}, m}}$, to be the same as in S16:
\begin{align} \label{eq:scaling_euclid}
        \ln (M_\mathrm{bias}/M_0) = B_{M,0} + \alpha_\sfont{E} \ \ln(1+z)\,, \\    
        \sigma^2_{M_{{\rm obs}, m}} =  \sigma^2_{M_{\rm obs},0} -1 +(1+z)^{2\,\beta}\,.  
\end{align}
To estimate the number counts, we considered equally spaced redshift bins ranged in $z \in [0.9, 2.0]$, with a width of $\Delta z=0.1$. As for the limiting mass, we defined a mass selection function similar to the one in S16. In our analysis, we performed two types of tests. In the first one, we fixed the nuisance parameters above to the best fit of $(B_{M,0},\, \alpha_\sfont{E},\,  \sigma_{M_{\rm obs},0},\, \beta)=(0.0 \pm 0.05,\, 0.0 \pm 0.05,\, 0.2 \pm 0.07,\, 0.125 \pm 0.00625)$, while in the second, we found the best-fit values for the same parameters, as is described in \Cref{sec:results}.  

\subsubsection{\textit{Rubin}-like cluster number counts forecast}

For \textit{Rubin}, we followed  \citet{2018arXiv180901669T} and \citet{2018ApJ...854..120M}, which modeled the galaxy cluster number counts as a function of their observed cluster
richness. The probability of observing a certain richness, $\lambda$, given a mass, $M$, was modeled as
\begin{equation}
    P(\ln \lambda|M)\,\d \ln \lambda \equiv \frac{1}{\sqrt{2\pi}\, \sigma_{\ln \lambda |M}}\, \exp \Bigg[-\frac{x^2(\lambda,M)}{2 \, \sigma^2_ { \lambda|M } } \Bigg]\, \d\ln \lambda \,,
\end{equation}
with $x$ defined by  
\begin{equation}
    x(\lambda,M) \equiv \ln \lambda - \Bigg[A + B \, \ln \Bigg(\frac{M}{M_\mathrm{pivot}} \Bigg) + C \ln \left( 1+z \right) \Bigg]\,,
    \label{eq:scalingLSST}
\end{equation}
with $M_\mathrm{pivot} = 3 \times 10^{14} \,h^{-1} M_\odot$, and $A$, $B$, and $C$ the dimensionless constants set below.
Thus, the number counts at a given redshift and richness bin can be described by the following expression:
\begin{align}
    N = & \int_{\Omega_{\rm tot}} \d \Omega \int^{z_{\rm max}}_{z_{\rm min}} \d z\; \frac{\d V}{\d z\,\d\Omega}
    \int \d M \;n(M,z) 
    \int^{\lambda_{\rm   max}}_{\lambda_{\rm min}} \frac{\d\lambda}{\lambda}\;P(\ln \lambda | M) \nonumber \\
    = & \int_{\Omega_{\rm tot}} \d \Omega \int^{z_{\rm max}}_{z_{\rm min}}\d z \; \frac{\d V}{\d z\,\d\Omega} \int \d M\; n(M,z) 
    \,S(M|\lambda_{\rm min}, \lambda_{\rm max})\, ,
\end{align}
where the subscripts ${\rm min}$ and ${\rm max}$ denote the minimum and maximum in each bin and $S(M|\lambda_{\rm min}, \lambda_{\rm max})$ is expressed as
\begin{equation}
\begin{split}
     S(M|\lambda_{\rm min}, \lambda_{\rm max}) \equiv \int ^{\lambda_{\rm max}}_{\lambda_{\rm min}} \d\ln{\lambda}\; P(\ln{\lambda|M}) = \\ 
     = \frac{1}{2}\,\Bigg[ \text{erf}\, \Bigg(\frac{x(\lambda_{\rm max},M)}{\sqrt{2 \, \sigma_{ \lambda |M}}}\Bigg) - \text{erf}\, \Bigg(\frac{x(\lambda_{\rm min},M)}{\sqrt{2 \, \sigma_{ \lambda |M}}}\Bigg) \Bigg]\,. 
\end{split}
\end{equation}
erf is the error function, and the scatter, $\sigma_{\ln \varLambda|M}$, is expressed as
\begin{equation}
    \sigma_{ \lambda |M} = \sigma_0 + q_M \ln \Bigg(\frac{M}{M_\mathrm{pivot}} \Bigg) + q_z \ln \left( 1+z \right)\,.
\end{equation}
Here, we considered nine equally spaced redshift bins in the interval $[0.0, 0.9 ]$, and 20 richness bins within the interval $[20,220]$ with a sky coverage of $\Omega_{\rm tot} = 18\,000\,{\rm deg}^2$.
 In \Cref{sec:results} we show results for the $\Lambda$LTB forecast, obtained by fixing the parameters $(\sigma_0,\,A,\, B,\, C,\, q_M,\, q_z)$ to the best fit of $(0.456, 25 \pm 5,\, 1 \pm 0.8,\, 0 \pm 1.2,\, 0 \pm 0.05,\, 0 \pm 0.2)$, as well as results when finding the best-fit values for these same parameters.

\subsection{{\rm eROSITA}-like survey X-ray-based cluster counts}\label{subSec:erosita}

Unlike the \Euclid and \textit{Rubin}
surveys, eROSITA detects X-ray emission from clusters. Thus, our scaling relation is based on X-ray properties, as is described in \citet{2012MNRAS.422...44P}.  
Galaxy clusters were sorted in terms of the photon counts that will be detected by the eROSITA telescope. The counts were converted to dark matter haloes by taking into account the properties of the X-ray detector and the integration
time of the observations. 
The X-ray cluster counts can be computed in a given redshift bin and observed temperature, $T$, as
\begin{align}
    N =  &\int_{\Omega_{\rm tot}} \d \Omega \int_{z_{\rm min}}^{z_{\rm max}} \d z \frac{\d V}{\d z\,\d\Omega}  \int \d M \; n(M,z)  \nonumber \\ 
    &  \hspace{2cm}       
    \times\int_{T(M_{\rm min})} \d\ln T_{\rm X}\, P(\ln T_{\rm X} | M)\,,
\end{align}
where the subscripts ${\rm min}$ and ${\rm max}$ denote the minimum and maximum in each redshift bin and $T_{X}=T/T_0$, where $T_0 = 1 \, {\rm keV}$ and $T(M_{\rm min})$ is the temperature bin lower limit obtained using Eq.~\eqref{eq:scaling_eRosita_2}. This would correspond to the mass taken from the following range, $\logten (M_{\rm min}/M_0) \in [13.0,\, 13.7,\, 14.0]$, at the three redshift bins with edges $z \in [0.24,\, 0.3,\, 0.4,\, 0.52]$, and a sky coverage of $\Omega_{\rm tot} = 27\,500\,{\rm deg}^2$ for this survey.
$P(\ln T_{\rm X} | M)$ is the probability distribution of the X-ray temperature, $T_{\rm X}$, given the mass,
\begin{equation}
    P(\ln T_{\rm X}|M) = \frac{1}{\sqrt{2 \pi \, \sigma^2_{TM}}} \exp \Bigg\{ - \frac{\left[\ln T_{\rm X} - \mu_T\right]^2}{2 \, \sigma^2_{TM}} \Bigg\}\,,
    \label{eq:scaling_eRosita_2}
\end{equation}
$\mu_T$ is given by the scaling relation
\begin{eqnarray}
     \mu_T = \alpha_{TM} \ln\Bigg(\frac{M_{500}}{\beta_{TM}}\Bigg)\, + \, \beta_{TM} \ln E(z) \, 
    + \, \ln\Bigg(\frac{5\,{\rm keV}}{T_0}\Bigg)\,,%
    \label{eq:scaling_eRosita_1}
\end{eqnarray}
and $E(z)= H(z)/H_0$.
The best-fit parameter values for this relation and the scatter, $\sigma_{TM}$, are \citep[see][]{2012MNRAS.422...44P}
\begin{eqnarray}
    \alpha_{TM} &=& 0.65 \pm 0.03\,, \\ 
    \beta_{TM} &=& (3.02 \pm 0.11) \times 10^{14}\, h^{-1}\, M_\odot\,, \\ \sigma_{TM} &=& 0.119\,.
\end{eqnarray}

\subsection{{\rm DESI}-like survey velocity mass determination-based cluster counts}\label{subSec:desi}

In DESI, the total luminosity of a given cluster within a 1Mpc radius ($L_\mathrm{1Mpc}$) is the mass proxy. This luminosity is related (and calibrated) to the mass, according to the scaling relation 
Eq.~\eqref{eq:scaling_desi}, below, as in \citet{2021ApJS..253...56Z}, where a fast clustering algorithm to identify the clusters was applied to the photometric redshift catalog. The total masses of the galaxy clusters were derived using a calibrated richness–mass relation that is based on the observations of X-ray emission, and the Sunyaev-Zeldovich effect:
\begin{equation}
    \logten(L_\mathrm{1Mpc}/L^*) = \frac{\logten(M_{500}/M_0)}{a} - \frac{b}{a}\,\logten(1+z) - \frac{c}{a}\,,
    \label{eq:scaling_desi}
\end{equation}
where $L^*$ is the characteristic luminosity and $M_0 = 1 \, h^{-1}\, M_\odot$. 
The best-fit parameters of the above relation are \citep{2021ApJS..253...56Z}
\begin{equation}
    a= 0.81 \pm 0.02\,,\; b= 0.50 \pm 0.14\,,\; c = 12.61 \pm 0.04\,.
\end{equation}
The above parameters were obtained from a photometric catalog containing sources that are used to obtain the spectroscopic measurements of their redshift. 
Therefore, we rescaled the error of the redshift parameter, $b$, to reflect the improvements in the error of $z$ with respect to the expected accuracy of the spectroscopic measurements in \cite{DESI:2016igz} to reach more realistic constraints, and not heavily underestimate the power of DESI.\\ 
Given the observables of this survey, we modeled the number counts in a given redshift and luminosity bin as 
\begin{equation}
   N = \int_{\Omega_{\rm tot}} \d \Omega \int_{z_\mathrm{min}}^{z_\mathrm{max}} \d z\; \frac{\d V}{\d z\,\d\Omega} \int \d M\; n(M,z) 
    \int_{y_{\rm min}} \d y \,P\left(y\, | M\right) \,,
\label{eq:N_desi}
\end{equation}
where the subscripts ${\rm min}$ and ${\rm max}$ denote the minimum and maximum in each redshift bin, where $y=\logten (L_\mathrm{1\,Mpc}/L^*)$, and where $P \left(y\, | M\right)$ has a probability that follows a Gaussian distribution similar to Eq.~\eqref{eq:scaling_eRosita_2}.
For DESI, we took redshift bins with a thickness equal to 0.1 in the range $z=[0.0,\, 1.0]$, 
a sky coverage of $\Omega_{\rm tot} = 20\,000\,{\rm deg}^2$, 
and lower limits for $y$ as presented below,
for the numerical evaluation of the integrals in Eq.~\eqref{eq:N_desi}:
\begin{equation}
\begin{split}
    y_{\rm min}=\logten (L_\mathrm{1Mpc}/L^*)_{\rm min} \in [1.65,\, 2.13,\, 2.35,\, 2.50,\, 2.61,\\
      2.69, \, 2.76,\, 2.83,\, 2.88,\, 2.93]\, .
\end{split}
\end{equation}

\begin{figure}[]
\centering
\includegraphics[width=\columnwidth]{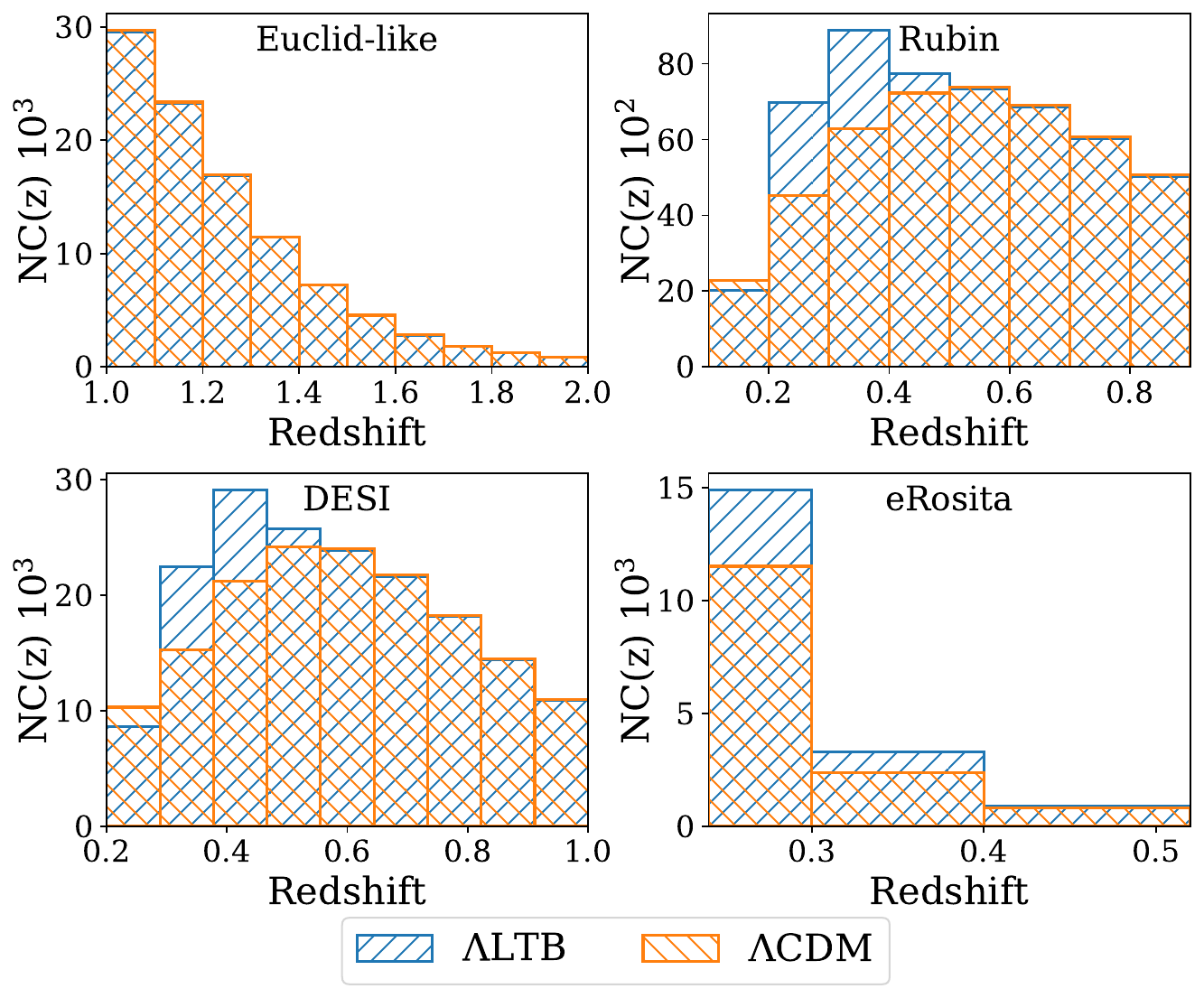}
\caption{
Bars with heights representing cluster number counts in each of the redshift bins considered for the different surveys used in this work for $\Lambda$CDM. $\{\delta_0, z_{\rm B}\} = \{0.0,0.0\}$ versus $\Lambda$LTB with $\{\delta_0,z_{\rm B}\} = \{-0.5,0.5\}$, while the other parameters were kept at their fiducial values. 
}
\label{fig:NCfourprobes}
\end{figure}

Finally, for illustrative purposes, Fig.~\ref{fig:NCfourprobes} shows integrated cluster counts in the redshift bins of the surveys used in this work, for a given $\Lambda$LTB model versus our fiducial $\Lambda$CDM cosmology. The cluster counts modifications within the $\Lambda$LTB model are, as expected, correlated with the void profile chosen in this work. In particular, we see that the \Euclid sample change to the cluster counts is not noticeable. This is because its redshift bins fall outside the LTB void in the values adopted here. Though it will still serve to break degeneracies with the FRW cosmological parameters, \Euclid is also expected to observe lower redshifts than our sample; however, here we already used those redshifts for the \textit{Rubin} sample, since we considered both as our baseline unified survey. Other estimators were used in \cite{Euclid:2022ucc}, as was detailed in the introduction, with similar constraints on the cluster counts, as we shall see later, but we still need to model other observables within $\Lambda$LTB, such as galaxy lensing or clustering, in order to be able to further benefit from \Euclid or other similar high-redshift surveys for which the aforementioned probes constitute their main observational targets.

In the following sections, we consider three forecast scenarios that differ by the degree of control one may have on the different systematic effects related to the nuisance parameters involving the scaling relations of the surveys described in this section. In \Cref{tab:baseline_parameters} we list the cosmological and nuisance parameters of our modeling as well as the priors assumed in each scenario. These scenarios are:
\begin{itemize}
\item Optimistic settings: only the cosmological parameters are left free to vary with flat priors. In this case the nuisance parameters are fixed to their fiducial values.
\item Realistic settings: the cosmological parameters are left free, while we consider Gaussian priors for the nuisance parameters of our probes.
\item Pessimistic settings: the cosmological as well as the nuisance parameters of all surveys are left free to vary.
\end{itemize}

\section{Results} \label{sec:results}
\renewcommand{\arraystretch}{1.7}
\begin{table}[]
\caption{Cosmological and nuisance parameter priors. The fiducial values are at the center of the display ranges, except for the fiducial value of $z_\sfont{B} $, which was set to $z_B=0$.}
\centering
\resizebox{\columnwidth}{!}{%
\begin{tabular}{l  c c c c c} 
\hline
\multirow{2}{*}{cosmo. param.} & {$\delta_0$}   & {$z_\sfont{B} $} & {$\Omega_{\rm m,0} $} & {$h $} & {$\sigma_8$}\\
 & {[}$-$1, 1{]} & {[}0, 0.5{]} & {[}0.1, 0.9{]} & {[}0.3, 1.2{]} & {[}0.7, 1.0{]}   \\ \hline
\multirow{2}{*}{$Euclid$} & $B_{M,0}$ & $\alpha_\sfont{E}$ & $\sigma_{M_{\rm obs},0}$ & $\beta$ &  \\
 & {[}$-$0.05, 0.05{]} & {[}$-$0.05, 0.05{]} & {[}0.13, 0.27{]} & {[}0.124, 0.126{]} &  \\ \hline
\multirow{2}{*}{Rubin} & $A$ & $B$ & $C$ & $q_M$ & $q_z$ \\
 & {[}20, 30{]} & {[}0.2, 1.8{]} & {[}$-$1.2, 1.2{]} & {[}$-$0.05, 0.05{]} & {[}$-$0.2, 0.2{]} \\ \hline
\multirow{2}{*}{eRosita} & $\alpha_{TM}$ & $\beta_{TM}/10^{14} \, h^{-1}\,M_\odot$ &  &  &  \\
 & {[}0.62, 0.68{]} & {[}0.291, 3.13{]} &  &  &  \\ \hline
\multirow{2}{*}{DESI} & $a$ & $b$ & $c$ &  &  \\
 & {[}0.79, 0.83{]} & {[}0.4995, 0.5005{]} & {[}12.57, 12.65{]} &  &  \\ \hline
\end{tabular}
}
\label{tab:baseline_parameters}
\end{table}
\renewcommand{\arraystretch}{1}

We are mainly interested in assessing, as closely as possible, the constraints on the $\Lambda$LTB model from a \Euclid-like survey when optimally combined with \textit{Rubin}-like survey data.
For that, our baseline approach was to combine cluster observations from \Euclid and \textit{Rubin}, knowing that \Euclid is expected to perform better at high redshifts, for $z > 0.9$, whereas \textit{Rubin} provides information for $z < 0.9$. We started with an optimistic case, assuming that we would have good control over the systematic effects, and thus we fixed 
their associated nuisance parameters to their fiducial values. We also fixed the correction parameter, $\alpha$, in \Cref{eq:delta_alpha} to $\alpha = 2$, the value found by simulations to give the best fit \citep{2012PDU.....1...24A}. Furthermore, we also explored the case where $\alpha$ is left free to show that the impact on the results is minimal, as is seen in Fig. \ref{fig:free_alpha} of \Cref{Sect:alphacase}. Finally, as in  \cite{Euclid:2022ucc}, we chose a flat prior for $z_{\rm B} \in [0, 0.5]$, since for $\delta_0\,\approx\,0$ the parameter space is highly degenerate with higher arbitrarily values of $z_{\rm B}$. This baseline was then compared to the case in which we added data from eROSITA or DESI, these two being independent of each other. Thus, this led to three additional cases of combining survey information: \Euclid + \textit{Rubin} + DESI, \Euclid + \textit{Rubin} + eROSITA, and finally \Euclid + \textit{Rubin} + DESI + eROSITA. We also show the results of combining all the probes. These combinations are expected to offer tighter constraints than our baseline since we also fixed the nuisance parameters for these two probes in our first case.

We then repeated the same approach, but this time allowed the nuisance parameters of the baseline probes to vary within large priors in less optimistic configurations. We also combined \Euclid and \textit{Rubin} datasets with the eROSITA and DESI datasets in this scenario in order to break the degeneracies and balance the expected weakening of the constraints in the baseline case with \Euclid and \textit{Rubin} alone. This is expected to tighten the constraints even with wide priors on the primary probes' nuisance parameters. Finally, we ran a further case, in which we combined all four probes at once, but this time with the most pessimistic settings, leaving all the nuisance parameters free to vary, in order to assess the loss of precision from this full auto-calibration approach.\\   
\renewcommand{\arraystretch}{1.2}
\begin{table*}[]
\caption{Relative differences of the 95\% confidence level with respect to the baseline (\Euclid + \textit{Rubin}) + DESI + eROSITA pessimistic case. Here we auto-calibrated the nuisance parameters, i.e., we assumed a non-informative prior for those parameters, as is described in \Cref{Sec:ResultsPessimistic}, while in the optimistic case we assumed fixed nuisance parameters, and in the \enquote{prio.} case we adopted a Gaussian prior for the nuisance parameters, taking into account the values, and their reported uncertainty as the width of the Gaussian, as is mentioned in \Cref{sec:probes}. A positive value indicates a gain, while the negative sign indicates a loss. }
\centering
\resizebox{2\columnwidth}{!}{%
\begin{tabular}{c c c c c c c c c}
 &
  Base. opt. &
  \begin{tabular}[c]{@{}c@{}}Base.+\\ DESI opt.\end{tabular} &
  \begin{tabular}[c]{@{}c@{}}Base.+\\ eROSITA opt.\end{tabular} &
  \begin{tabular}[c]{@{}c@{}}Base.+DESI\\+eROSITA opt.\end{tabular} &
  Base. prio. &
  \begin{tabular}[c]{@{}c@{}}Base.+\\ DESI prio.\end{tabular} &
  \begin{tabular}[c]{@{}c@{}}Base.+\\ eROSITA prio.\end{tabular} &
  \begin{tabular}[c]{@{}c@{}}Base.+DESI\\ +eROSITA prio.\end{tabular} \\ \hline
\multicolumn{1} {c}     {$\delta_0      $}      &       3.5\%   &       4.6\%   &       4.1\%   &       4.1\%   &       3.5\%   &       2.3\%   &       4.1\%   & 3.5\%   \\      \hline
\multicolumn{1} {c}     {$z_\sfont{B}   $}      &       21.9\%  &       25.9\%  &       28.3\%  &       61.9\%  &       $-$1.4\%        &       28.3\%  &       15.3\%  & 38.0\%  \\      \hline
\multicolumn{1} {c}     {$h     $}      &       87\%    &       93.0\%  &       88.9\%  &       93.8\%  &       $-$76.6\%       &       53.2\%  &       25.5\%  & 58.5\%  \\      \hline
\multicolumn{1} {c}     {$\sigma_8      $}      &       71.9\%  &       80.5\%  &       81.9\%  &       87.1\%  &       $-$461.9\%      &       41.4\%  &       25.7\%  & 60\%    \\      \hline
\multicolumn{1} {c}     {$\Omega_{\rm m,0}      $}      &       47.5\%  &       71\%    &       81\%    &       84\%    &       $-$275\%        &       50\%    &       -5.0\%  & 61.5\%  \\      \hline
\end{tabular}%
}
\label{tab:parameter_values}
\end{table*}
\renewcommand{\arraystretch}{1}
\renewcommand{\arraystretch}{1.2}

\begin{table}[]
\caption{Relative difference of the 95\% confidence levels with respect to \citet{Euclid:2022ucc}, for the optimistic case in our analysis, in which we fixed our nuisance parameters, as is described in \Cref{Sec:ResultsOptimistic}. To be on an equal footing with that work, we show the relative difference in terms of the amplitude of the primordial power spectrum of scalar perturbations, $A_
{\rm s}$, because the aforementioned work did not output constraints for $\sigma_8$. A positive value indicates a gain, while the negative sign indicates a loss.}
\centering
\resizebox{0.60\columnwidth}{!}{%
    \begin{tabular}{c c}
    &
    \begin{tabular}[c]{@{}c@{}}Base.+DESI +eROSITA opt.\end{tabular} \\ \hline
\multicolumn{1}{c} {$\delta_0   $}      &       $-$3.1\%                \\      \hline
\multicolumn{1}{c} {$z_\sfont{B}        $}          &   36.3\%     \\   \hline
\multicolumn{1}{c} {$h $}          &    55.4\%      \\  \hline
\multicolumn{1}{c} {$A_{\rm s}$}        &   53.3\%      \\ \hline
\multicolumn{1}{c} {$\Omega_{\rm m,0} $}   &    64.0\%     \\ \hline
\end{tabular}%
}
\label{tab:parameters_wrtdavid}
\end{table}
\renewcommand{\arraystretch}{1}

\subsection{Optimistic case assuming good control over the systematic effects}\label{Sec:ResultsOptimistic}

\begin{figure}[]
\centering
\includegraphics[width=\columnwidth]{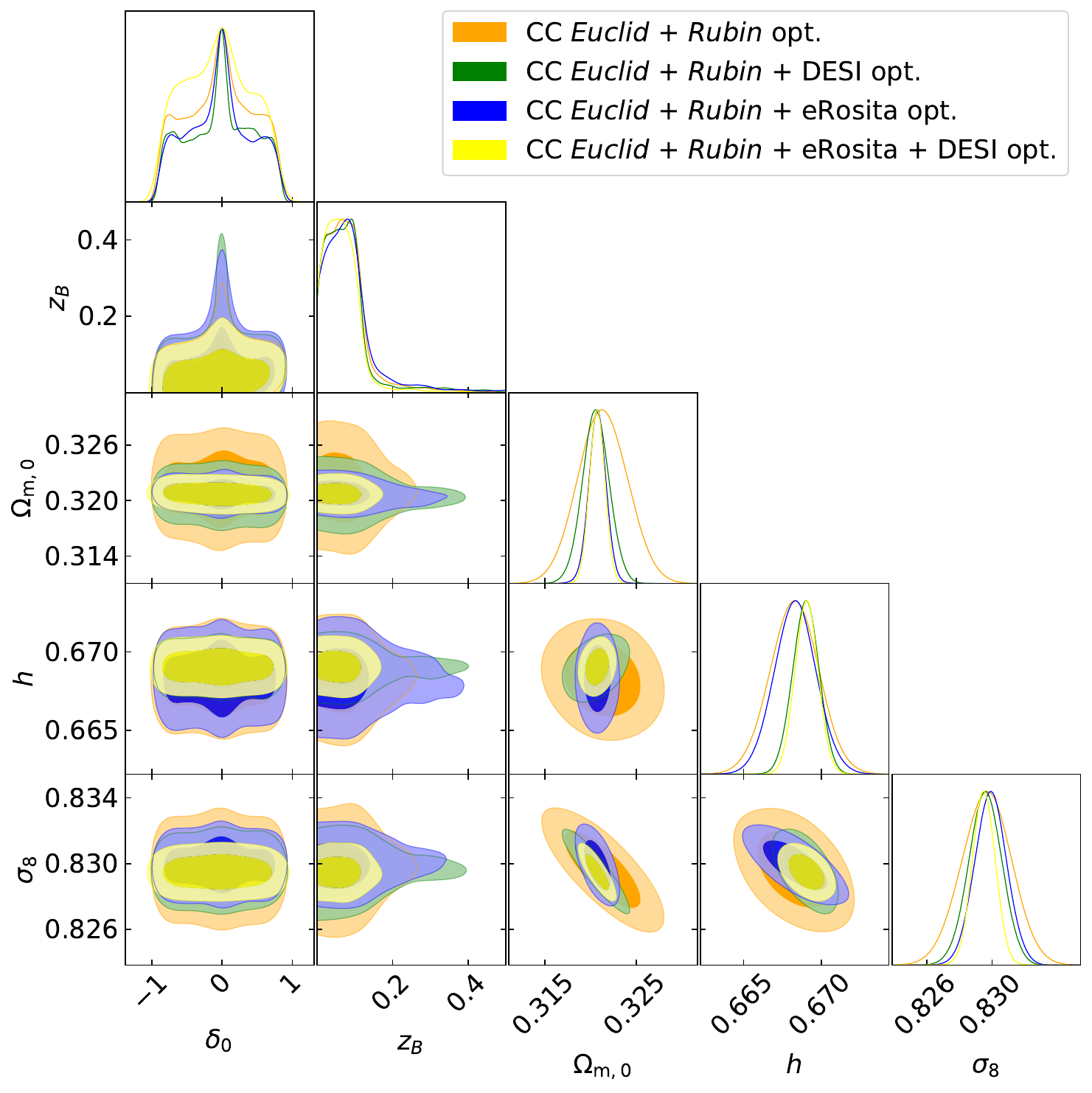}
\caption{
$\Lambda$LTB 68\% (darker color) and 95\% (lighter color) confidence contours for \Euclid and \textit{Rubin} survey probes (orange), for when we include the DESI survey probe (green), the eROSITA survey probe (blue), and both (yellow) in the optimistic case (see subsection~\ref{Sec:ResultsOptimistic}). 
}
\label{fig:cosmo_constrains_opt}
\end{figure}

In this subsection, we start by showing, in Fig. \ref{fig:cosmo_constrains_opt}, the marginalization of the parameters considered for the forecast using our baseline (\Euclid + \textit{Rubin}). We also show, in the same figure, the baseline with the addition of the secondary survey probes described in subsections \ref{subSec:erosita} and \ref{subSec:desi}, all with fixed nuisance parameters, as this will be our optimistic case, in which we have control over the systematic effects. We also show in \Cref{tab:parameter_values} the gain in accuracy for all the combinations and cases, which we discuss later with respect to the most pessimistic case in which all the parameters, including the nuisance ones, were left free to vary.

We first observe that the cosmological parameters, $\Omega_{\rm m,0}$, $h$, and $\sigma_8$, are constrained to a percent level while the density of the void, $\delta_0$, does not prefer any value over the null one corresponding to its FLRW value. However, the redshift border, $z_{\rm B}$, of the void is constrained to 0.4 and below at the 95\% confidence level. 
Furthermore, as one could expect, when we add additional information from the secondary probes, we better constrain the FLRW-like parameters, especially $\Omega_{\rm m,0}$, with the latter showing a gain of about 40\%, as is seen in \Cref{tab:parameter_values}, while $h$ and $\sigma_8$ increase by more than $10\%$. For the $\Lambda$LTB parameters, $\delta_0$ still shows no preference within each optimistic case, while the $z_{\rm B}$ constraints increase by a factor of three when we combine all of the probes. We also notice that the baseline plus DESI offers similar constraints as the baseline plus eROSITA - due to the fact that eROSITA covers a larger sky area than DESI, while the latter spans a wider range and higher redshifts. Additionally, as was expected, our gain is further improved when we combine all of the probes. Finally, we also compare our results with the work of \cite{Euclid:2022ucc}, as is shown in \Cref{tab:parameters_wrtdavid}, where we only compare the optimistic case with all of the probes, so as to be on an equal footing with the aforementioned author's work. Overall, we get better constraints on our parameters than \cite{Euclid:2022ucc}, getting an improvement of about 50\% for the FLRW parameters, 35\% for $z_\sfont{B}$, and almost the same for $\delta_0$.

\subsection{Realistic case assuming Gaussian priors on the nuisance parameters}\label{Sec:ResultsRealistic}

Here we discuss the case in which we consider nuisance parameters with Gaussian priors, with averages and variances taken from those given in \Cref{tab:baseline_parameters}. In this more realistic case, we consider ourselves to have less control over the systematic effects compared to the previous scenario, and therefore we allow some scattering in the nuisance parameters, thereby expecting looser constraints with respect to the optimistic case. On the other hand, when we combine or add more probes we should get tighter constraints on our parameters in any of the settings considered.\\

\begin{figure}[]
\centering
\includegraphics[width=1\columnwidth]{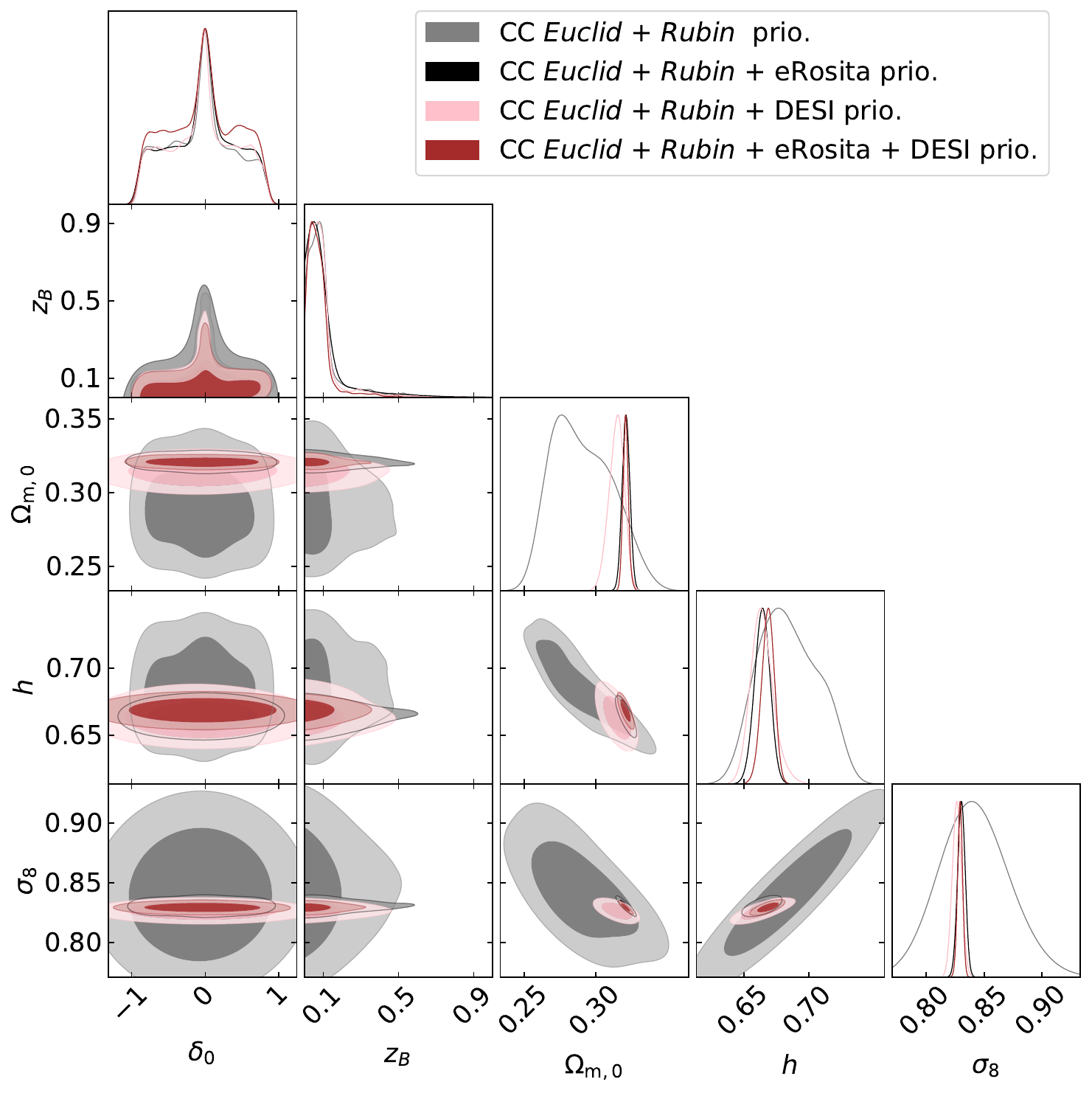}
\caption{$\Lambda$LTB and nuisance 68\% (darker color) and 95\% (lighter color) confidence contours for our realistic case, in which we considered Gaussian priors for the nuisance parameters in \Cref{tab:baseline_parameters} in the calibration of the scaling relations. Here, we show the forecasts for the baseline (gray), and we also include secondary probes in the contours (\Euclid + \textit{Rubin} + DESI (black), \Euclid + \textit{Rubin} + eROSITA (pink), and finally \Euclid + \textit{Rubin} + DESi + eROSITA (red)).}
\label{fig:cosmo_constrains_prio}
\end{figure}

In Fig. \ref{fig:cosmo_constrains_prio} we show the contours within this scenario for the three cases: the baseline plus DESI, the baseline plus eROSITA, and the baseline plus DESI and eROSITA, while the gains with respect to the most pessimistic case are also in the same \Cref{tab:parameter_values}, along with those for the optimistic scenario.
 As predicted, the optimistic case shows better constraints when compared to its realistic counterpart, close to the same gain we observe when using cluster counts to constrain FLRW cosmological parameters. For the LTB parameters, we observe an overall improvement in the constraints, albeit smaller than that for the FLRW parameters. This is also supported by the general gain seen in the detailed \Cref{tab:parameter_values}, where the constraining power is degraded in most cases by more than a factor of two for each analogous probe, with the greatest decrease found when we compare the baseline only, reaching a factor of three for $\Omega_{\rm m,0}$ and $\sigma_8$. The difference lessens when we choose to add the two additional surveys, DESI or eROSITA, to the baseline, and the lowest value is reached when we simultaneously add these two probes, staying around 20\% lower on average. This can also be seen in Fig. \ref{fig:cosmo_constrains_prio_pess}, where we plot the two different analyses (the optimistic case is plotted in yellow and the realistic case in red) of the chains that include all the considered probes. 

\subsection{Pessimistic constraints with auto-calibration of the nuisance parameters}\label{Sec:ResultsPessimistic}

\begin{figure}[]
\centering
\includegraphics[width=1\columnwidth]{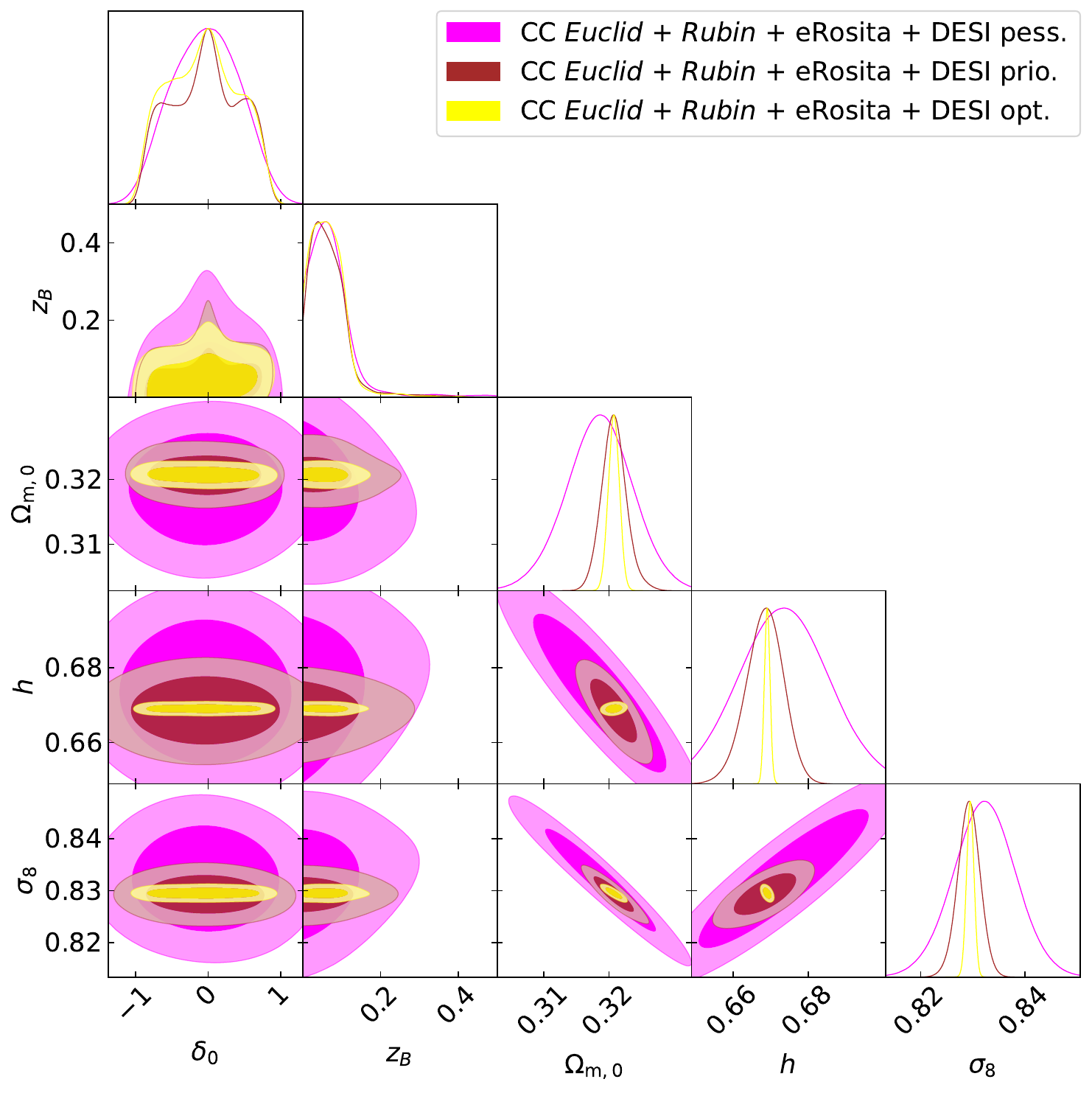}
\caption{ 
$\Lambda$LTB 68\% (darker color) and 95\% (lighter color) confidence contours considering all the probes in this paper (\Euclid + \textit{Rubin} + eROSITA + DESI). Here, we compare the optimistic case (yellow) \ref{Sec:ResultsOptimistic} with the realistic one (red) \ref{Sec:ResultsRealistic} and finally the pessimistic one (fuchsia) \ref{Sec:ResultsPessimistic}.}
\label{fig:cosmo_constrains_prio_pess}
\end{figure}

Finally, we present the most \enquote{pessimistic} case, in which we consider all the nuisance parameters of our probes to be free, by using non-informative flat priors for our Bayesian analysis. In Fig.~\ref{fig:cosmo_constrains_prio_pess}, we show the cosmological and LTB model parameters forecast for this scenario, together with the analogous chains for the realistic and optimistic scenarios. The contours have widened with respect to the more optimistic ones, since most of the gains in \Cref{tab:parameter_values} are positive; however, even when considering non-informative priors for the nuisance parameters, we are still able to put constrains on the LTB parameters at a higher level than in the baseline realistic case. This is seen in \Cref{tab:parameter_values}, which shows the power of a combination of probes to break and auto-calibrate the degeneracies between the $\Lambda$LTB and the nuisance parameters.

\section{Conclusions} \label{sec:conclusions}

In this work we have forecast the forthcoming constraints for $\Lambda$LTB, an inhomogeneous model, using galaxy cluster abundance mock catalogs as our only probe, from `Stage-IV'-like experiments such as \Euclid and \textit{Rubin}, as our baseline. We have also studied the impact of taking into account observables from other experiments such as DESI and eROSITA.

We assumed different scenarios, reflecting our degree of confidence in the collected data: an optimistic case in which we fixed the nuisance parameters of the galaxy cluster scaling laws, a realistic case in which we allowed free scaling parameters within the Gaussian prior uncertainties, and finally a pessimistic case in which we assumed non-informative priors on the nuisance parameters.

We find that our combination of probes yields stronger constraints, especially for the non-LTB parameters. The constraints are degraded when the scaling laws are allowed more freedom in their parameter space. However, the combination of all probes is still able to break the degeneracies and auto-calibrate all of the nuisance parameters, even when they are left free to vary.

We also find that the $\Lambda$LTB model is still viable, even if the data turn out to be compatible with $\Lambda$CDM in the future. This was also the case in \cite{Euclid:2022ucc}, which combined a plethora of geometrical probes as well as probes relating to the growth of structures, without including clusters. However, since our probe is further tackling the nonlinear scales for structure formation, while retaining some geometrical information from the sky coverage of the different surveys, we are able to get similar or better constraints for some parameters. Moreover, we highlight that we have separated the redshift domain of \Euclid and \textit{Rubin} (for $0<z<1$ we consider \textit{Rubin} only, while for \Euclid we consider the redshift range of $1<z<2$). While such an analysis was not performed in \cite{Euclid:2022ucc}, our conservative approach is still competitive, highlighting the importance of the galaxy cluster counts probe in constraining cosmology within the $\Lambda$LTB model.



\begin{acknowledgements}
The authors would like to thank Stefano Camera for useful comments and style checks on the draft. ZS acknowledges funding from DFG project 456622116 and support from the IRAP and IN2P3 Lyon computing centers.
This work was financed by Portuguese funds through FCT - Funda\c c\~ao para a Ci\^encia e a Tecnologia in the framework of the following Research Projects: EXPL/FIS-AST/1368/2021 (AdS, AC, JPM), PTDC/FIS-AST/0054/2021 (NN, AdS, AC, JPM), UIDB/04434/2020, UIDP/04434/2020, CERN/FIS-PAR/0037/2019 (NN, AdS, AC, JPM), and 2022.04048.PTDC (CJM). CJM also acknowledges FCT and POCH/FSE (EC) support through Investigador FCT Contract 2021.01214.CEECIND/CP1658/CT0001, and AC acknowledges support form the FCT research grant 2020.06644.BD. JGB and SN acknowledge support from the Research Project PID2021-123012NB-C43 [MICINN-FEDER], and the Centro de Excelencia Severo Ochoa Program CEX2020-001007-S. DS acknowledges financial support from
the Fondecyt Regular project number 1200171. D. C. would also like to thank the Robert E. Young Origins of the Universe Chair fund for its generous support. This work was financed by FEDER – Fundo Europeu de Desenvolvimento Regional – funds through the COMPETE 2020 – Operational Programme for Competitiveness and Internationalisation (POCI) – and by Portuguese funds of FCT, under projects PTDC/FIS-AST/0054/2021 and UIDB/04434/2020 \& UIDP/04434/2020.

\end{acknowledgements}

\bibliographystyle{aa}
\bibliography{biblio,Euclid}


\clearpage

\begin{appendix}
\section{Case with a free nonlinear correction prescription parameter}
\label{Sect:alphacase}

\begin{figure}[]
\centering
\includegraphics[width=\columnwidth]{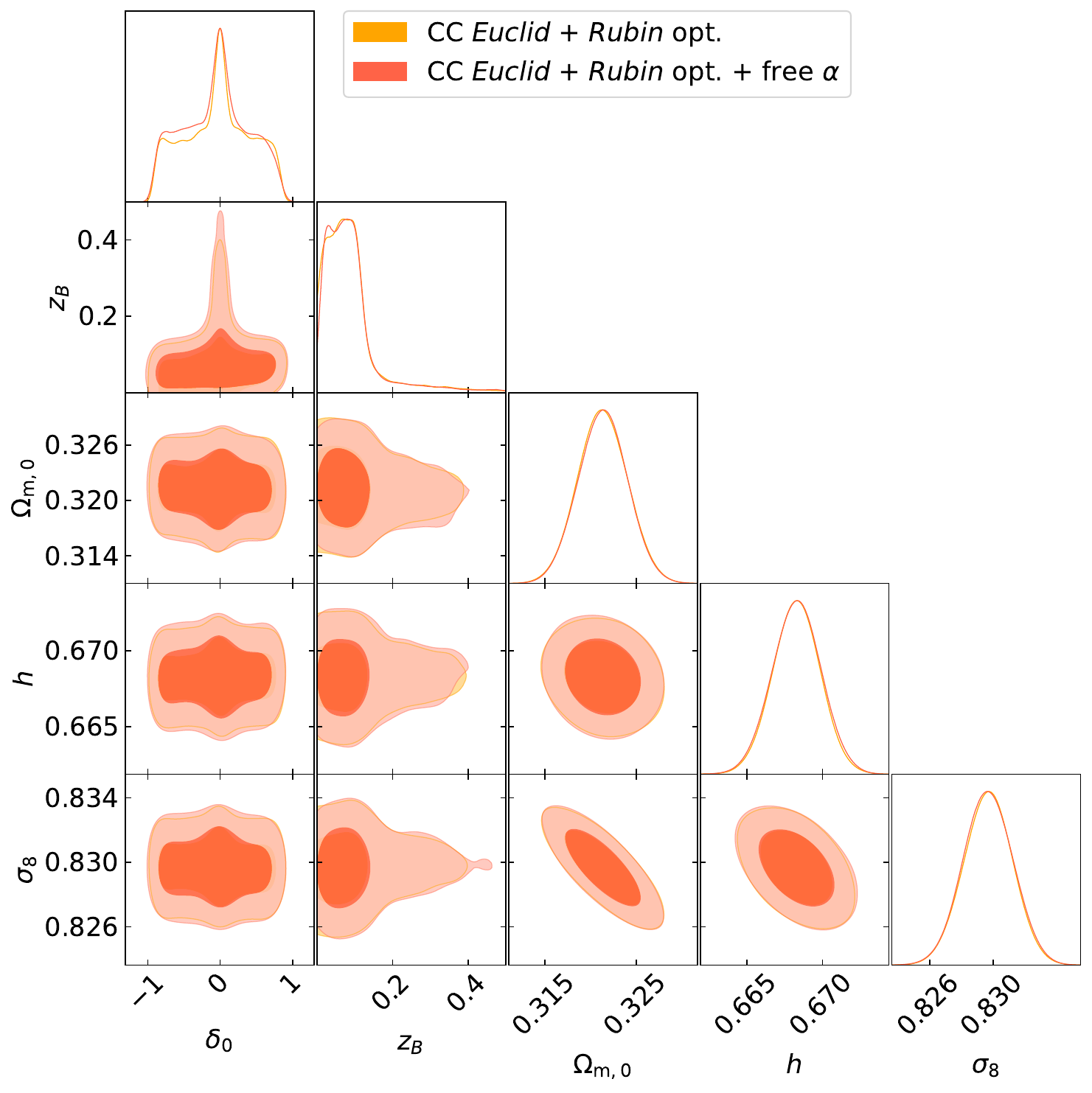}
\caption{
$\Lambda$LTB 68\% (darker color) and 95\% (lighter color) confidence contours for the baseline of our analysis: the \Euclid and \textit{Rubin} survey probes. In yellow we present the case in which we model $\alpha$ to a fixed value, whereas in red we display the case where our modeling of the problem includes a free $\alpha$ with a wide prior, $\in [0,4]$.}
\label{fig:free_alpha}
\end{figure}

As is mentioned in the main text, here we show the optimistic case of our baseline (\Euclid + \textit{Rubin}) in Fig.~\ref{fig:free_alpha}, where we compare the constraints obtained from having a fixed $\alpha = 2$ (Eq. \ref{eq:delta_alpha}) to when we set this parameter as free to vary. We observe that the contours of these two cases are similar, as was expected from the simulations in \citet{2010PhRvD..82l3530A} justifying fixing $\alpha$ for the other case's analysis.

\end{appendix}

\end{document}